\documentclass[lettersize,journal]{IEEEtran}
\usepackage{amsmath,amsfonts}
\usepackage{amsmath}
\usepackage{amssymb}
\usepackage{algorithmic}
\usepackage{algorithm}
\usepackage{array}
\usepackage[caption=false,font=normalsize,labelfont=sf,textfont=sf]{subfig}
\usepackage{textcomp}
\usepackage{stfloats}
\usepackage{url}
\usepackage{verbatim}
\usepackage{graphicx}
\usepackage{cite}
\usepackage{hyperref}
\usepackage{booktabs}
\usepackage{bm}
\begin{document}

\title{Global Frequency Reference Tracking as an Oscillation Suppression Mechanism in VSM Primary Control: A Coupled-Oscillator Study}
\author{Taha Saeed Khan,~\IEEEmembership{Student Member,~IEEE,}
\thanks{Taha Saeed Khan is with the School of Electrical Computer Engineering, Oklahoma State University, Stillwater 74075, OK, USA}
\thanks}

\markboth{}
{Shell \MakeLowercase{\textit{et al.}}: A Sample Article Using IEEEtran.cls for IEEE Journals}

\maketitle

\begin{abstract}
Synchronization in power systems is traditionally achieved through physical network coupling, whereby inverter-based resources (IBRs) and synchronous machines converge to a common frequency via oscillatory swing dynamics. In conventional operation, secondary control acts on a slow time scale and is typically engaged only after the primary dynamics have largely settled. As a result, in the absence of an explicit global reference, disturbances can induce prolonged transients and large phase excursions.

This work considers a setting in which the total active power balance is known and maintained at all times, and proposes a control architecture for virtual synchronous machine (VSM) based inverters in which all units track a broadcast global frequency reference. Under this assumption, synchronization is transformed from a mutual oscillator-locking problem into a reference-tracking problem.

Using a second-order swing-network model, we show that embedding a simple proportional–integral (PI) frequency controller can significantly improves transient behavior. A washout mechanism ensures that the additional control action vanishes in steady state, thereby preserving network-determined power sharing. Simulations on a three-oscillator network demonstrate reduced frequency overshoot, elimination of underdamped oscillations, and lower angular stress compared to conventional open-loop synchronization, highlighting the effectiveness of a global frequency reference as a coordination mechanism for grid-forming inverter networks.
\end{abstract}

\section{Introduction}

The synchronization of generator clusters in power systems has traditionally been analyzed through absolute rotor-angle and frequency dynamics\cite{Preece2024}. In both synchronous-machine-dominated grids and modern inverter-dominated systems, frequency emerges as a collective variable governed by the balance between power injections, network coupling, inertia, and damping. In the absence of an explicit global reference, synchronization is achieved implicitly through physical interactions among generators, leading to oscillatory transient behavior following disturbances \cite{Xu2007}.

When analyzed in a stationary coordinate system, any disturbance such as during faults, generator tripping, or load changes, frequency restoration relies on large rotor-angle excursions and prolonged inter-area oscillations \cite{Lachs1987}, which impose significant stress on machines, converters, and network components.

Unlike synchronous machines, a cluster of Grid forming (GFM) IBRs are capable of establishing voltage and frequency autonomously, but they still rely predominantly on local power–frequency feedback and physical coupling through the network to achieve synchronization. Consequently, even with advanced control structures, frequency regulation remains fundamentally an oscillatory synchronization problem which experience similar challenges as faced by conventional synchronous machines \cite{Fan2023}.

Although inverter-based resources (IBRs), particularly grid-forming inverters, are inherently more capable than synchronous machines in autonomously establishing voltage and frequency, they still predominantly rely on conventional generator-like mechanisms—namely local power–frequency feedback and physical network coupling—for synchronization. Consequently, even with advanced GFM control architectures, frequency regulation remains fundamentally an oscillatory synchronization process unless augmented by an explicit reference or secondary coordination layer \cite{Guo}.

In this work, we consider a setting in which the aggregate active power balance is perfectly known, i.e., total generation equals total consumption at all times. In modern power systems, this assumption is increasingly realistic \cite{Huang} due to high-resolution measurement, fast communication, and active energy management systems that can estimate system-wide power demand and adjust the active power setpoints of inverter-based resources (IBRs) within sub-second duration. Under this assumption, global frequency drift caused by net power imbalance is eliminated, allowing the analysis to focus exclusively on synchronization and transient dynamics. Building on this premise, a control architecture is proposed in which all grid-forming resources track a broadcast global frequency reference. Hence, by embedding the system dynamics in a synchronous rotating reference frame, the synchronization problem is transformed from mutual oscillator locking into a reference-tracking task.

This change in perspective has important consequences. Rather than relying on network-induced oscillations to redistribute power, frequency deviations can be directly regulated through simple modified primary control actions. As shown in this paper, augmenting standard grid-forming dynamics with a proportional–integral (PI) controller acting on frequency deviation enables fast, non-oscillatory convergence to equilibrium. A washout mechanism \cite{Asadi2015} is employed to ensure that the control action vanishes at steady state, thereby preserving the natural power-flow solution and network-determined power sharing.

The remainder of this paper formalizes these ideas using a second-order Kuramoto (swing-network) model, which provides a compact and physically interpretable abstraction of VSM based inverter synchronization. Through numerical simulations, it is demonstrated that how the introduction of a global frequency reference fundamentally can alter the transient behavior of grid-forming networks, yielding improved damping, reduced frequency overshoot, and significantly smaller angular excursions.

\section{ Grid-Forming (GFM) Inverters Frequency Control: Three Canonical Laws}

In practical grid-forming inverters, the frequency variable is generated internally and evolves according to a chosen grid-forming law. The three most common families are: (i) droop-based grid forming \cite{alghamdi2021}, (ii) virtual synchronous machine (VSM)\cite{Xin}, and (iii) oscillator-based methods such as dispatchable virtual oscillator control (dVOC)  \cite{Gross} or matching-control-type oscillators. Below equations provides a representative summary for each class.

\subsubsection {Droop-Based Grid-Forming Control}
Droop control implements an algebraic power--frequency relation:
\begin{equation}
\omega_i(t) = \omega_0 - m_{p,i}\big(P_i(t)-P_i^\star\big),
\label{eq:droop_omega}
\end{equation}
where $\omega_0$ is the nominal angular frequency, $m_{p,i}>0$ is the active-power droop coefficient, $P_i(t)$ is the measured active power injection, and $P_i^\star$ is the active power setpoint.
Equivalently, the phase angle is obtained by integration:
\begin{equation}
\dot{\theta}_i(t) = \omega_i(t).
\label{eq:droop_theta}
\end{equation}
\noindent
\textit{Key Property:} Droop generally permits a nonzero steady-state frequency deviation when $P_i(t)\neq P_i^\star$.

\subsubsection{ Virtual Synchronous Machine (VSM) / Swing-Emulation}
VSM control embeds frequency in a first-principles-like inertial law:
\begin{equation}
M_i \dot{\omega}_i(t) + D_i\big(\omega_i(t)-\omega_0\big)
= P_i^\star - P_i(t),
\label{eq:vsm_swing}
\end{equation}
\begin{equation}
\dot{\theta}_i(t) = \omega_i(t),
\label{eq:vsm_theta}
\end{equation}
where $M_i>0$ is a virtual inertia constant and $D_i>0$ is a damping coefficient.

\noindent
\textit{Key Property:} Frequency dynamics are inertial (second-order in $\theta_i$), improving transient behavior, but a secondary layer is needed to converge to reference frequency.

\subsubsection{Oscillator-Based Grid Forming (e.g., dVOC / Matching-Type)}
Oscillator-based GFM methods generate an internally stable limit cycle and synchronize through feedback. A generic complex-oscillator form can be written as:

\begin{small}
\begin{equation}
\dot{v}_{\alpha\beta,i}(t) =
\omega_0 J v_{\alpha\beta,i}(t)
+ \kappa_i\Big(v_{\alpha\beta,i}^\star(t) - v_{\alpha\beta,i}(t)\Big)
+ \Psi_i\big(P_i(t),Q_i(t)\big),
\label{eq:oscillator_generic}
\end{equation}
\end{small}
where $v_{\alpha\beta,i}\in\mathbb{R}^2$ is the voltage vector in the stationary $\alpha\beta$ frame, $J=\begin{bmatrix}0 & -1\\ 1 & 0\end{bmatrix}$ is the $90^\circ$ rotation matrix, $\kappa_i>0$ is a synchronization gain, $v_{\alpha\beta,i}^\star$ is a voltage reference (magnitude/phase objective), and $\Psi_i(\cdot)$ denotes the power-based feedback that enforces desired $P$-$Q$ behavior.
One may also express the oscillator in polar coordinates $v_{\alpha\beta,i}=V_i[\cos\theta_i,\ \sin\theta_i]^\top$, yielding an implicit frequency evolution through $\dot{\theta}_i$.

\noindent
\textit{Key Property:} Frequency emerges from the oscillator dynamics and power-based feedback.. This is the only control where reference frequecy appears in primary control.

As explicit in the case of VSM, the applied control law inherently permit frequency deviations following disturbances, unless a secondary control mechanism, analogous to Automatic Generation Control (AGC) in synchronous generating units is activated.

However, IBRs control operates on significantly faster time scales and does not necessarily require a distinct secondary control layer. Instead, the frequency-restoration functionality can be embedded directly within the primary grid-forming controller. This integration can significantly improve transient performance and mitigate oscillatory behavior.

To explicitly drive frequency deviations to zero, the primary grid-forming control law can be augmented with a frequency-tracking loop:

This paper demonstrates that incorporating reference frequency tracking within the primary control layer is a key enabling mechanism for improving frequency regulation using simple Proportional--Integral (PI) control laws.

\section{Generator as $2^{nd}$ order Kuramoto Oscillator}

The classical swing equation provides the foundational dynamic model emulated by
VSM-based controllers for power-system synchronization. For a conventional
synchronous generator, the dynamics can be written as follows.

Let $\delta_i$ denote the electrical rotor angle of generator $i$, and define the
synchronous reference frame $\theta_i := \delta_i - \omega_s t$, where $\omega_s$
is the nominal synchronous speed. Then $\dot{\theta}_i = \omega_i - \omega_s$.

The classical swing equation is
\begin{equation}
M_i \dot{\omega}_i + D_i(\omega_i-\omega_s) = P_{m,i} - P_{e,i},
\label{eq:swing}
\end{equation}
where $M_i$ and $D_i$ denote inertia and damping, and $P_{m,i}$ and $P_{e,i}$ are
mechanical and electrical powers, respectively. The swing equation states that the rotor accelerates or decelerates according to the imbalance between mechanical input power and electrical output power, filtered by inertia and damping.

Under a balanced positive-sequence network model, the electrical power injection
can be written as
\begin{equation}
P_{e,i} = \sum_{j=1}^N V_iV_j\!\left(G_{ij}\cos(\theta_i-\theta_j)
+ B_{ij}\sin(\theta_i-\theta_j)\right).
\label{eq:Pe_full}
\end{equation}
Assuming a lossless (or weakly resistive) network ($G_{ij}\approx 0$) and
approximately constant voltage magnitudes, \eqref{eq:Pe_full} reduces to the
sinusoidal coupling form
\begin{equation}
P_{e,i} \approx \sum_{j=1}^N K_{ij}\sin(\theta_i-\theta_j),
\qquad K_{ij}:=V_iV_jB_{ij}.
\label{eq:Pe_kuramoto}
\end{equation}

Substituting \eqref{eq:Pe_kuramoto} into \eqref{eq:swing} and expressing the
dynamics in the synchronous frame yields the second-order Kuramoto (swing-network) model
\begin{equation}
M_i \ddot{\theta}_i + D_i \dot{\theta}_i
= P_{m,i} - \sum_{j=1}^N K_{ij}\sin(\theta_i-\theta_j).
\label{eq:second_order_kuramoto}
\end{equation}

\section{Three-Generator Inertial Kuramoto}

\subsection{Swing-network dynamics}
Three coupled generators using the inertial Kuramoto (swing-network)
model in a synchronous rotating frame are simulated. Let $\theta_i(t)$ denote the phase angle
of generator $i$ and $\omega_i(t)=\dot{\theta}_i(t)$ denote its frequency deviation.
The dynamics are:
\begin{align}
\dot{\theta}_i &= \omega_i, \label{eq:sim_dtheta}\\
M_i \dot{\omega}_i + D_i \omega_i
&=
P_i(t) - \sum_{j=1}^N K_{ij}\sin(\theta_i-\theta_j), i=1,\dots,N,
\label{eq:sim_swing}
\end{align}
where $M_i>0$ and $D_i>0$ represent inertia and damping, respectively.
\begin{equation}
K_{ij} =
\begin{cases}
K_0, & i\neq j,\\
0, & i=j.
\end{cases}
\label{eq:sim_coupling}
\end{equation}

\subsection{Step disturbance with power balance}

A step disturbance is applied to the mechanical power inputs at time $t_0$:
\begin{equation}
P(t) = P_{\mathrm{base}} + u(t-t_0)\,P_{\mathrm{step}},
\label{eq:sim_step}
\end{equation}
where $u(\cdot)$ is the unit-step function. 

To ensure that the disturbance excites only relative (inter-area) dynamics and does not induce a spurious acceleration of the system's center-of-inertia (COI) mode, the mechanical power inputs are constructed to be power-balanced at all times.

First, a baseline mechanical power vector $\bm{P}_{\text{base}} \in \mathbb{R}^N$ is selected such that
\begin{equation}
    \bm{1}^\top \bm{P}_{\text{base}} = \sum_{i=1}^N P_{\text{base},i} = 0,
\end{equation}
which guarantees that, prior to the disturbance, the net mechanical torque applied to the system is zero. Consequently, the synchronous (COI) frequency remains constant and no global acceleration is present.

Next, a step disturbance vector $\bm{P}_{\text{step}} \in \mathbb{R}^N$ is constructed by applying a power increment to a selected generator and then removing its mean value, enforcing
\begin{equation}
    \bm{1}^\top \bm{P}_{\text{step}} = \sum_{i=1}^N P_{\text{step},i} = 0.
\end{equation}

This normalization ensures that the step disturbance redistributes power among generators without altering the total mechanical input. As a result, the disturbance excites only the relative swing modes of the system rather than the COI mode. Without this constraint, the system would experience a net torque imbalance, leading to a uniform drift of all frequencies and obscuring synchronization behavior.


All simulations are performed in a synchronous rotating reference frame; hence, $\omega_i$ denotes frequency deviation from the nominal value, and $\omega_i=0$ corresponds to operation at the rated 60 Hz frequency.

\subsection{Equilibrium-consistent initial conditions}
To initiate the simulation from a synchronized steady-state operating point (a ``flat start'' in frequency), we set the initial frequency deviations to zero:
\begin{equation}
    \omega(0) = \mathbf{0}.
\end{equation}
The initial phase angles $\theta(0) = \theta_0$ are computed as the steady-state solution to the open-loop power-balance equations under the baseline mechanical power $P_{\text{base}}$:
\begin{equation}
    P_{\text{base},i} - \sum_{j=1}^N K_{ij} \sin(\theta_{0,i} - \theta_{0,j}) = 0, \quad i = 1, \dots, N.
\end{equation}
In the numerical implementation, this nonlinear system of equations is solved using the \texttt{fsolve} algorithm in Matlab \cite{Matlab} to ensure that $\dot{\omega}(0) = 0$, thereby preventing initial spurious transients prior to the applied disturbance.

\subsection{Center of inertia as relative reference}
To remove the arbitrary global phase reference, relative angles are computed via the COI angle

\begin{equation}
\theta_{\mathrm{COI}}(t) := \frac{\sum_{i=1}^N M_i\theta_i(t)}{\sum_{i=1}^N M_i},
\qquad
\theta_i^{\mathrm{rel}}(t) := \theta_i(t) - \theta_{\mathrm{COI}}(t).
\label{eq:sim_coi}
\end{equation}

\section{Three-Generator Inertial Kuramoto with Reference-Tracking PI}

An explicit global frequency setpoint $\omega_{\mathrm{ref}}$ is used in this case where again a network of three inertial Kuramoto is simulated.
Let $\phi_i(t)$ denote the phase angle of generator $i$ in this rotating frame, and let
$\Delta\omega_i(t)$ denote the frequency deviation from $\omega_{\mathrm{ref}}$.
This converts synchronization from an open-loop ``mutual locking'' problem into a \emph{tracking} problem where $\Delta\omega_i \to 0$ is directly enforced .

\subsection{Closed-loop swing-network dynamics}
The inertial Kuramoto (swing-network) model with all-to-all coupling is
\begin{align}
\dot{\phi}_i &= \Delta\omega_i, \qquad i=1,\dots,N, \label{eq:dphi}\\
M_i \dot{\Delta\omega}_i + D_i \Delta\omega_i
&= P_i(t) - \sum_{j=1}^{N} K_{ij}\sin(\phi_i-\phi_j) + u_i(t), \label{eq:ddw}
\end{align}

Compared to the open-loop inertial Kuramoto model \eqref{eq:dphi}--\eqref{eq:ddw} without $u_i$,
the reference-tracking controller introduces \emph{additional tuning knobs} $(K_{p,i},K_{i,i})$ that can
accelerate convergence to $\Delta\omega_i=0$ without modifying the physical parameters $(M_i,D_i)$.

\subsection{Reference-tracking PI control law}
A PI controller on frequency deviation is now implemented as:
\begin{align}
u_i(t) &= -K_{p,i}\Delta\omega_i(t) - K_{i,i} z_i(t), \label{eq:ui}\\
\dot{z}_i(t) &= \Delta\omega_i(t), \label{eq:zi}
\end{align}
where $K_{p,i}\ge 0$ provides fast proportional ``frequency pull'' and $K_{i,i}\ge 0$ eliminates steady-state frequency error via integral action.
This structure is consistent with the notion that a globally shared reference allows $K_p$ and $K_i$ to act as active damping/stiffness knobs.

\subsection{Step disturbance with power balance}
Same step disturbance is applied to the mechanical power inputs at time $t_0$:
\begin{equation}
P(t) = P_{\mathrm{base}} + \mathbf{1}_{t\ge t_0}\, P_{\mathrm{step}},
\label{eq:Pstep}
\end{equation}
where $\mathbf{1}_{t\ge t_0}$ is the unit-step indicator.

Again power balance is enforced:
\begin{equation}
\mathbf{1}^\top P_{\mathrm{base}} = 0, \qquad \mathbf{1}^\top P_{\mathrm{step}} = 0,
\label{eq:powerbalance}
\end{equation}
where the second constraint is implemented by removing the mean of the injected step vector.

\subsection{Equilibrium-consistent initial conditions}
To initiate the simulation from a synchronized operating point (a ``flat start'' in frequency), we set the initial frequency deviation and controller state vectors to zero:
\begin{equation}
    \bm{\Delta\omega}(0) = \mathbf{0}, \quad \bm{z}(0) = \mathbf{0}.
\end{equation}
The initial phase angle vector $\bm{\phi}(0) = \bm{\phi}_0$ is computed as the steady-state solution to the power-balance equations under the baseline mechanical power $\bm{P}_{\text{base}}$:
\begin{equation}
    P_{\text{base},i} - \sum_{j=1}^N K_{ij} \sin(\phi_{0,i} - \phi_{0,j}) = 0, \quad i = 1, \dots, N.
\end{equation}
This system is solved numerically using the \texttt{fsolve} algorithm to ensure that $\bm{\dot{\Delta\omega}}(0) = \mathbf{0}$, thereby isolating the system's pure transient response to the subsequent step disturbance.


\subsection{Reference-tracking PI control law with Washout Filter}

To ensure that the control effort $u_i(t)$ provides transient damping without altering the long-term steady-state power-flow equilibrium, a washout filter is applied to the integrator state:
\begin{align}
    u_i(t) &= -K_{p,i}\Delta\omega_i(t) - K_{i,i}z_i(t), \label{eq:washout_u} \\
    \dot{z}_i(t) &= \Delta\omega_i(t) - \frac{1}{\tau}z_i(t), \label{eq:washout_z}
\end{align}
where $\tau > 0$ represents the washout time constant.

\subsection{Steady-State Consistency}

In the steady state, the system reaches a synchronized frequency where $\Delta\omega_i = 0$ and $\dot{z}_i = 0$. Substituting these conditions into \eqref{eq:washout_z} yields $z_i = 0$, which consequently forces the control torque to vanish:
\begin{equation}
    \lim_{t \to \infty} u_i(t) = 0. \label{eq:u_limit}
\end{equation}
As $u_i(t)$ converges to zero, the closed-loop swing equation reduces to the natural power-balance requirement:
\begin{equation}
    P_i(t) - \sum_{j=1}^N K_{ij} \sin(\phi_i - \phi_j) = 0. \label{eq:final_sharing}
\end{equation}
This confirms that while the controller eliminates oscillatory transients, the final power sharing is determined solely by the physical network topology.

\section{Simulation Results and Discussion}

The natural open-loop response against the closed-loop reference-tracking controller equipped with the washout filter is now simulated and compared. The paramentes used for the simulations are provided in Table \ref{tab:sim_params}.

\begin{table}[h]
\centering
\caption{Simulation Parameters for the 3-Generator Inertial Kuramoto Network}
\label{tab:sim_params}
\begin{tabular}{lll}
\hline
\textbf{Category} & \textbf{Parameter} & \textbf{Value} \\
\hline
Network &
Number of generators $N$ & 3 \\

Dynamics &
Inertia constants $M_i$ & $[2.0,\; 3.0,\; 2.5]$ \\
&
Damping coefficients $D_i$ & $[3.0,\; 3.0,\; 3.0]$ \\

Coupling &
Coupling gain $K_0$ & $8.0$ \\
&
Coupling matrix $K_{ij}$ & $K_0(1-\delta_{ij})$ \\

Control &
Proportional gains $K_{p,i}$ & $[8.0,\; 4.0,\; 3.0]$ \\
&
Integral gains $K_{i,i}$ & $[4.0,\; 2.0,\; 1.0]$ \\
&
Washout time constant $\tau$ & $1.0~\text{s}$ \\

Disturbance &
Baseline mechanical power $P_{\text{base}}$ & $[0.6,\; -0.3,\; -0.3]$ \\
&
Step magnitude $\Delta P$ & $2.0$ \\
&
Step time $t_0$ & $3.0~\text{s}$ \\

Initialization &
Initial frequency deviation & $\Delta\omega_i(0)=0$ \\
&
Initial washout state & $z_i(0)=0$ \\
&
Initial angles & Solved via \texttt{fsolve} \\
\hline
\end{tabular}
\end{table}

\begin{figure}[htbp]
\centering
\includegraphics[width=0.4\textwidth]{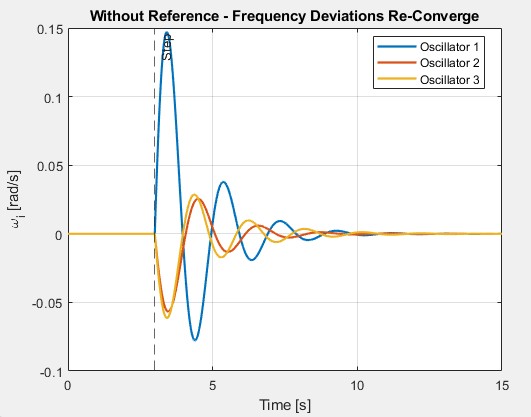}
\caption{Transient frequency deviations of a three-oscillator network without global frequency reference tracking. After a step disturbance, frequency restoration relies exclusively on network coupling, leading to pronounced oscillatory behavior.
}
\label{fig:1A}
\end{figure}

\begin{figure}[htbp]
\centering
\includegraphics[width=0.4\textwidth]{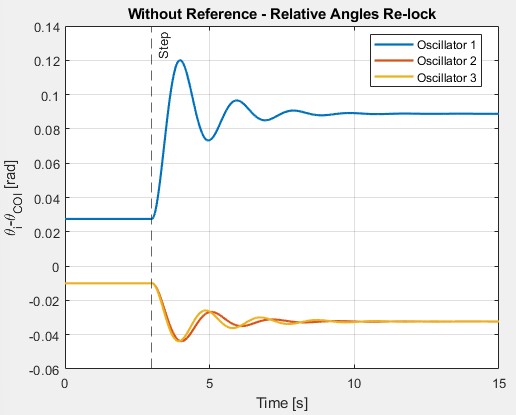}
\caption{Relative phase angles of the three oscillators without a global frequency reference. Following the disturbance, the oscillators re-lock through network coupling, and the steady-state phase offsets encode the resulting power-sharing equilibrium.
}
\label{fig:1B}
\end{figure}

\begin{table}[htbp]
\centering
\caption{Steady-State Power Sharing Verification: Natural Case (No Reference)}
\label{tab:natural_power}
\begin{tabular}{lccc}
\toprule
\textbf{Osc.} &
\textbf{$P_m$ (pu)} &
\textbf{ $P_e$ (pu)} &
\textbf{Error} \\
\midrule
1 & 1.9333   & 1.9332   & $1.0\times10^{-4}$ \\
2 & -0.96667 & -0.96698 & $3.1\times10^{-4}$ \\
3 & -0.96667 & -0.96624 & $-4.3\times10^{-4}$ \\
\midrule
\textbf{Total Sum} &
\textbf{0.00000} &
\textbf{0.00000} &
--- \\
\bottomrule
\end{tabular}
\end{table}

\begin{figure}[htbp]
\centering
\includegraphics[width=0.4\textwidth]{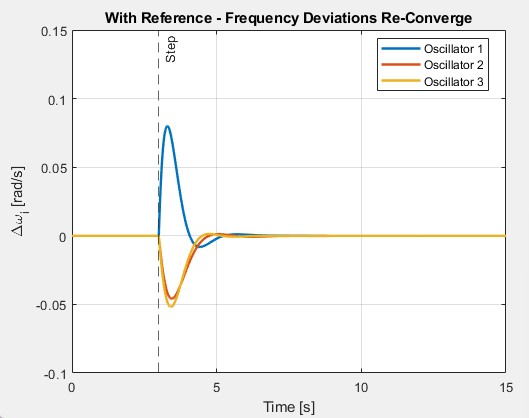}
\caption{Frequency deviations with global reference-frequency tracking. Following the disturbance, all oscillators rapidly re-converge to the reference frequency with strongly damped transients.}
\label{fig:2A}
\end{figure}

\begin{figure}[htbp]
\centering
\includegraphics[width=0.4\textwidth]{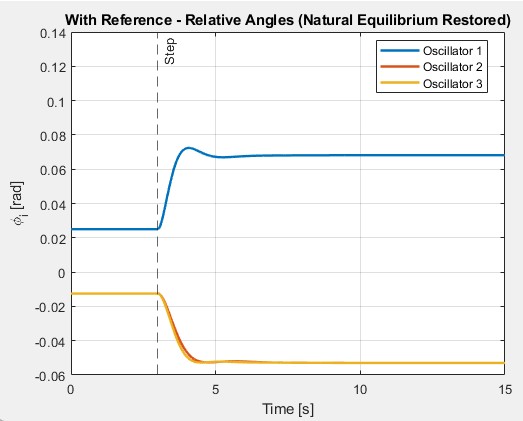}
\caption{With reference. Relative phase angles of the three oscillators under global reference-frequency tracking. After the disturbance, the washout mechanism restores the natural equilibrium, and the steady-state phase offsets encode the network-determined power sharing.
}
\label{fig:2B}
\end{figure}

\begin{table}[htbp]
\centering
\caption{Steady-State Power Sharing Verification: Reference-Tracking Controller Case}
\label{tab:ref_power}
\begin{tabular}{lcccc}
\toprule
\textbf{Osc.} &
\textbf{$P_m$ (pu)} &
\textbf{Control effort $u$ (pu)} &
\textbf{$P_e$ (pu)} &
\textbf{Error} \\
\midrule
1 & 1.9333   & $-5.92\times10^{-6}$ & 1.9333   & 0.0000 \\
2 & -0.96667 & $4.32\times10^{-6}$  & -0.96666 & 0.0000 \\
3 & -0.96667 & $1.59\times10^{-6}$  & -0.96667 & 0.0000 \\
\bottomrule
\end{tabular}
\end{table}

\subsection{Transient Stability and Oscillation Suppression}
The frequency deviation dynamics under a 2.0~pu step disturbance are illustrated in Fig.~\ref{fig:1A} and Fig.~\ref{fig:2A}. 

\begin{itemize}
    \item \textbf{Overshoot Reduction}: In the natural case (Fig.~\ref{fig:1A}), the system exhibits a peak frequency deviation of approximately 0.15~rad/s. In contrast, the reference-tracking architecture (Fig.~\ref{fig:2A}) limits this overshoot to less than 0.08~rad/s, representing a nearly 50\% improvement in transient frequency stability.
    \item \textbf{Damping Performance}: The open-loop system relies solely on network coupling and inherent damping $D_i$, leading to prolonged oscillations that persist for over 10~seconds. The proposed controller provides active damping via the $K_p$ and $K_i$ knobs, pulling frequency deviations to zero within 5~seconds with virtually no secondary ringing.
\end{itemize}

\subsection{Phase Dynamics and Equilibrium Consistency}
The evolution of relative phase angles is shown in Fig.~\ref{fig:1B} and Fig.~\ref{fig:2B}. 

\begin{itemize}
    \item \textbf{Non-Oscillatory Locking}: The natural phase response is characterized by large underdamped swings before settling. The reference-tracking case exhibits a smooth, monotonic transition to the new synchronized state, which shows significant reduction in the stress on the converters and on the transmission lines possible.
    \item \textbf{Equilibrium Alignment}: Although the transient paths differ, the final angular separation between the leading oscillator (Osc. 1) and the lagging oscillators (Osc. 2, 3) is identical in both scenarios ($\approx 0.120$~rad).
\end{itemize}

\subsection{Numerical Validation of Power Sharing}
Numerical verification of the steady-state power balance is provided in Table~\ref{tab:natural_power} and Table~\ref{tab:ref_power}. Table~\ref{tab:ref_power} shows that the steady-state control effort $u$ is effectively eliminated by the washout mechanism, reaching negligible values on the order of $10^{-6}$~pu. As a result of $u \to 0$, the electrical power export $P_e$ matches the mechanical input $P_m$ exactly.

\section{Conclusion}

This paper studied oscillation suppression in grid-forming/VSM-type synchronization using a second-order Kuramoto (swing-network) model. In the conventional setting without an explicit global reference, frequency restoration and phase re-locking occur solely through network
coupling, which leads to underdamped transients and large angular excursions following
disturbances.

Assuming that the aggregate active-power balance is maintained, we proposed a primary-control augmentation in which all units track a broadcast global frequency reference via a simple PI loop acting on frequency deviation. The resulting closed-loop dynamics transform synchronization from mutual oscillator locking into a reference-tracking problem, yielding substantially improved
transient behavior. Simulation results on a three-oscillator network showed reduced frequency overshoot, faster settling, and the elimination of underdamped ringing compared to the natural
(open-loop) response.

\end{document}